\documentstyle[prl,aps,multicol,psfig]{revtex}

\begin{document}

\input epsf
\title{Time evolution of the extremely
diluted Blume-Emery-Griffiths neural network}
\author{D. Boll\'e$^1$, D. R. C. Dominguez$^2$, 
R. Erichsen Jr.$^3$\\
E. Korutcheva$^4$
{\thanks{Permanent Address: G.Nadjakov Inst. Solid State Physics, 
Bulgarian Academy of Sciences, 1784 Sofia, Bulgaria }} 
and W. K. Theumann$^3$}
\address{$^1$Instituut voor Theoretische Fysica, Katholieke Universiteit    
Leuven, B-3001 Leuven, Belgium\\
$^2$E.T.S. Inform\'atica, Universidad Aut\'onoma de Madrid, Cantoblanco,
28049 Madrid, Spain\\ 
$^3$Instituto de F{\'\i}sica, Universidade Federal do Rio Grande do Sul, 
Caixa Postal 15051. 91501-970 Porto Alegre, RS, Brazil\\
$^4$Departamento de F\'{\i}sica Fundamental, UNED, c/Senda del Rey No. 9,
28080, Madrid, Spain
}

\date{\today}
\maketitle
\thispagestyle{empty}

\begin{abstract}
The time evolution of the extremely diluted Blume-Emery-Griffiths neural
network model is studied, and a detailed equilibrium phase diagram is
obtained exhibiting pattern retrieval, fluctuation retrieval and  
self-sustained activity phases. It is shown that saddle-point solutions 
associated with fluctuation overlaps slow down considerably the flow of 
the network states towards the retrieval fixed points. A comparison 
of the performance with other three-state networks is also presented. 

\end{abstract}


\begin{multicols}{2}


\section{Introduction}

There has been much interest in understanding the properties and 
predicting the behavior of large attractor neural networks. Of 
primary concern are the storage capacity, the ability and quality of 
retrieval, and the information transmitted by the network \cite{HK91}. 
The study of attractor neural networks has often been guided by the 
search for networks of optimal performance. 

It has been argued recently that  mutual information is the most 
appropriate concept to measure the performance quality, especially in 
sparsely coded networks \cite{DB98,BA00}. An attempt has been made to 
infer the Hamiltonian, or energy function, of an optimally performing 
three-state network from the structure of the initial mutual information 
and a disordered Blume-Emery-Griffiths (BEG) (\cite{BEG,ACN00} and
references therein) network model 
has been obtained. This has been used to derive the specific properties 
that characterize the performance of an extremely diluted network 
\cite{DK00}. The arguments leading to the BEG Hamiltonian and the 
dynamical behavior in networks of other architecture have been studied 
recently \cite{BV02,BE02}.

A characteristic feature of the model is to store and retrieve patterns 
and some of their fluctuations giving rise, in the thermodynamic
limit, to two independent local random fields. In contrast to the usual 
three-state network \cite{Y89,BS94}, both fields are self-adjusting 
functions and the network does not need an externally adjustable 
threshold parameter to activate the neuron states.

One of the further interesting aspects of the model is the presence 
of an independent order parameter macroscopically characterizing these 
fluctuations that could yield a new information carrying phase. However, 
neither the time evolution of this order parameter nor the stability 
of such a phase have been studied before. The purpose of the present 
paper is precisely to investigate these points in the extremely diluted 
network,  which has an exactly solvable dynamics. This also allows one 
to figure out the size of the basins of attraction, a point that has 
not been emphasized before, and that can only be studied to some extent 
in the diluted network due to the complexity of the interactions in the 
underlying model.

The motivation for our work is the discovery of moderately and, 
eventually, very long transients in the dynamic evolution of some
of the states of the network and the recognition that these states often 
drive the network into a retrieval phase. It also turned out that these 
states are not stable in the absence of synaptic noise (temperature $T$), 
in contrast to an earlier claim \cite{DK00}, but that a finite, 
activity dependent, threshold value for $T$ is required for these states 
to stabilize. We also produce a further comparison of the performance 
of the network with other three-state networks. Preliminary results of 
this work have been presented recently \cite{DK02}.

The outline of the paper is the following. In Sec. II we summarize 
briefly the model and refer the interested reader to previous works for
further discussion \cite{DK00,BV02}. In Sec. III we state the 
macrodynamics and we present our results in Sec. IV, ending with our  
conclusions in Sec. V.

\section{The Model}

We consider a three-state network with symmetrically distributed neuron 
states $\sigma_{i,t}=0,\pm 1$ on sites $i=1,...,N$, at time step $t$, 
where $\sigma_{i,t}=\pm 1$ denote the active states. A set of $p$ ternary 
patterns, $\{\xi^{\mu}_{i}=0,\pm 1\}$, $\mu=1,...,p$, where 
$\xi_{i}^{\mu}= \pm 1$ are the active ones, are assumed 
to be independent random variables following the probability
distribution
\begin{equation}
p(\xi^{\mu}_{i})=
a\delta(|\xi^{\mu}_{i}|^{2}-1)
+(1-a)\delta(\xi^{\mu}_{i}) \, ,
\label{1}
\end{equation}
where the average $a=\left\langle(\xi_{i}^{\mu})^{2}\right\rangle$ 
is the activity of the patterns. 
These patterns are embedded in the network by means of a generalized 
learning rule together with a set $\{\eta_{i}^{\mu}\}$ 
of normalized fluctuations,
$\eta_{i}^{\mu}=((\xi_{i}^{\mu})^{2}-a)/a(1-a)$,  
of the binary patterns $(\xi_{i}^{\mu})^2$ about their average.

The learning rule consists of two Hebbian-like parts
\begin{equation}
J_{ij}= \frac{1}{a^2 N} 
\sum_{\mu=1}^{p}
\xi_{i}^{\mu} \xi_{j}^{\mu},\,\,\,\,\,\,\,\,\,\,\,\,\,\,
K_{ij}= \frac{1}{N} 
\sum_{\mu=1}^{p}
\eta_{i}^{\mu} \eta_{j}^{\mu} \, ,
\label{2}
\end{equation}
which are the random interactions in the bilinear BEG network  
with the dynamic variables $\sigma_{i,t}$ and  
$\sigma_{i,t}^2$, respectively \cite{DK00}. These are then used 
to construct the random local fields 
\begin{equation}
h_{i,t}=\sum_{j=1}^NJ_{ij}\sigma_{j,t},\,\,\,\,\,\,
\theta_{i,t}=\sum_{j=1}^N K_{ij}\sigma_{j,t}^2 \, ,
\label{3}
\end{equation}
where the first one is the usual local field for a three-state 
network. This enables one to obtain an effective single-site energy 
function for neuron $i$, in mean-field theory,
\begin{equation}
\epsilon_{i,t}=-[h_{i,t}\sigma_{i,t}+\theta_{i,t}\sigma_{i,t}^2]
\label{4}
\end{equation}\\
which rules the state-flip probability 
\begin{equation}
p(\sigma_{i,t+1}|\{\sigma_{i,t}\})=\exp(-\beta\epsilon_{i,t})/Z_t ,
\label{5}
\end{equation}
that specifies the parallel stochastic dynamics for the model, where
 $Z_t=1+ 2 e^{\beta\theta_t}\cosh(\beta h_t)$ and $\beta=a/T$ is the 
inverse temperature (noise) parameter.

In the sequel we do not indicate
the explicit $t$-dependence. In distinction to the usual 
three-state model \cite{Y89,BS94}, where the coefficient of the quadratic 
part in $\epsilon_{i}$ is an externally adjustable threshold parameter, 
we have here a self-adjusting, state and pattern dependent function
$\theta_{i}(\{\sigma_{j}\},\{\eta_{k}^{\mu}\})$. In this sense, the present 
model belongs to a wider class of ``self-control'' networks, a case of 
which has been discussed before \cite{DB98}. We remark, however, that the
threshold $\theta$ used in the latter is a macroscopic parameter, thus, no
average had to be done over the microscopic random variables at each time
step $t$.

Next, we consider the relevant quantities that describe the performance 
of the network. We need the conditional probability distribution of a 
neuron state given the distribution of the patterns,
$p(\sigma_{i}|\xi_{i}^{\mu})$. As a consequence of the mean-field
theory character of the model it is enough to consider the distribution
of a single typical neuron, so we can omit the index $i$. We can
then derive \cite{BD00}
\begin{equation}
p(\sigma|\xi^{\mu})=
(s_{\xi}+m^{\mu}\xi^{\mu}\sigma)\delta(\sigma^{2}-1)+
(1-s_{\xi})\delta(\sigma),
\label{6}
\end{equation} 
where
\begin{equation}
      s_{\xi}= s^{\mu}+l^{\mu}(\xi^{\mu})^{2},\,\,\,\,\, 
      s^{\mu}={q-an^{\mu}\over 1-a},\,\,\,\,\, 
      l^{\mu}={n^{\mu}-q\over 1-a}.
\label{7}
\end{equation}
Here, $m^{\mu}= \langle\langle\sigma
\rangle_{\sigma|\xi}\xi^{\mu}/a\rangle_{\xi}$
is the thermodynamic limit $N\rightarrow\infty$ of the retrieval overlap 
\begin{equation}
m_{N}^{\mu}={1\over aN}\sum_{i}\sigma_{i}\xi_{i}^{\mu}
\label{8}
\end{equation}
between the state of the network and pattern $\{\xi_{i}^{\mu}\}$, 
and the internal average is over the conditional probability.
The other parameters are the thermodynamic limits
$q=\langle\langle
\sigma^{2}\rangle_{\sigma|\xi}\rangle_{\xi}$ and
$n^{\mu}=\langle\langle
\sigma^{2}\rangle_{\sigma|\xi}(\xi^{\mu})^{2}/a\rangle_{\xi}$, 
of the neural activity
\begin{equation}
q_{N}=
{1\over N}\sum_{i}\sigma_{i}^{2}
\label{9}
\end{equation}
and the activity overlap \cite{BA00,DK00}
\begin{equation}
n^{\mu}_{N}= 
{1\over aN}\sum_{i}\sigma_{i}^{2}(\xi_{i}^{\mu})^{2},
\label{10}
\end{equation}
respectively. Finally, $l^{\mu}= \langle\langle
\sigma^2\rangle_{\sigma|\xi}{\eta^{\mu}}\rangle_{\xi}$, 
is the thermodynamic limit of the $fluctuation$ 
$overlap$ between the binary state variable $\sigma_{i}^2$ and 
$\eta_{i}^{\mu}$ defined as, 
\begin{equation}
l_{N}^{\mu}={1\over N}\sum_{i}\sigma_{i}^2\eta_{i}^{\mu}.
\label{11}
\end{equation}
As is clear from its definition, the fluctuation overlap is
connected  with the activity overlap.

An underlying assumption that leads to the BEG model
and that should be preserved in the implementation 
for any network architecture is that the dynamic activity 
$q\sim a$, as far as the order of magnitude is concerned. The
necessity of such an activity control system has been emphasized before
(cf. \cite{DB98,O96} and references therein).

The fluctuation overlap $l^{\mu}$ can be viewed as the 
retrieval overlap between the binary patterns $\{\eta_{i}^{\mu}\}$ 
and states $\{\sigma_{i}^{2}\}$. A priori, this will be 
independent of the retrieval overlap $m^{\mu}$ between the 
three-state patterns and states. As will be seen below, a finite 
non-zero retrieval overlap induces a finite fluctuation overlap, and 
in this case the parameter $l^{\mu}$ should not add anything essentially 
new to the three-state network. It will turn out, however, 
that the inverse is in general not true. Indeed,  
$l^{\mu}$ can be finite in a state of dynamic activity without 
necessarily a finite retrieval overlap $m^{\mu}$. Thus, there can be 
a phase in which $m^{\mu}=0$ but where, nevertheless, there is a finite
information carried by the fluctuation overlap $l^{\mu}\neq 0$. Whether 
the new phase is actually stable, interesting and how long it takes for 
the network to reach the corresponding states is the main issue that we 
address below.

Finally, we consider the mutual information between patterns and neurons, 
regarding the patterns as the inputs and the neuron states as the 
output of the network channel at each time step \cite{Sh48,Bl90}. This 
is given by, recalling that we do not indicate the explicit 
time dependence, 
\begin{equation}
I^{\mu}(\sigma,\xi^{\mu})
    =S(\sigma)-\langle S(\sigma|\xi^{\mu})\rangle_{\xi^{\mu}}, 
\label{13}
\end{equation}
where 
\begin{equation}
S(\sigma)=-q\ln(q/2)-(1-q)\ln(1-q)
\label{14}
\end{equation}
is the entropy and $S(\sigma|\xi^{\mu})\rangle _{\xi^{\mu}}=
aS_{a}+(1-a)S_{1-a}$
is the  equivocation term with
\begin{eqnarray}
S_{a}=-c_{+}^{\mu}\ln c_{+}^{\mu}-c_{-}^{\mu}\ln c_{-}^{\mu}
-(1-n^{\mu})\ln(1-n^{\mu})\nonumber\\
S_{1-a}=-s^{\mu}\ln(s^{\mu}/2)-(1-s^{\mu})\ln(1-s^{\mu}).
\label{15}
\end{eqnarray}
Here, $c_{\pm}^{\mu}=(n^{\mu} \pm m^{\mu})/2$ and $s^{\mu}$ is the 
parameter in the 
conditional probability $p(\sigma|\xi^{\mu})$. The mutual information can 
then be used to obtain the information $i^{\mu}=I^{\mu}\alpha$, where 
$\alpha=p/N$ is the storage ratio of the network.

\section{Macrodynamics}
The local fields in terms of the retrieval and the fluctuation overlaps 
\begin{equation}
h_{i}=\frac{1}{a}\sum_{\mu}\xi_{i}^{\mu}m^{\mu},\,\,\,\,\
\theta_{i}=\sum_{\mu}\eta_{i}^{\mu}l^{\mu},
\label{16}
\end{equation}
are the suitable starting point for the macrodynamics
\cite{DK00}. The actual overlaps
\begin{equation}
m^{\mu}= \langle\overline{\sigma_{i}}
{\xi_{i}^{\mu}\over a}\rangle_{\xi},\,\,\,\,\
l^{\mu}= \langle
\overline{\sigma_{i}^2}{\eta_{i}^{\mu}}\rangle_{\xi},
\label{17}
\end{equation}
depend on the thermal averages $\overline{\sigma_{i}}$ 
and $\overline{\sigma_{i}^2}$ given by 
$F_{\beta}(h,\theta)=
2e^{\beta\theta}\sinh(\beta h)/Z$ and 
$G_{\beta}(h,\theta)=2e^{\beta\theta}\cosh(\beta h)/Z$,
respectively. In the zero temperature limit
\begin{equation}
F_{\infty}=sgn(h)\Theta(|h|+\theta),\,\,\,\,\,\,\
G_{\infty}=\Theta(|h|+\theta).
\label{18}
\end{equation}

We assume that a single pattern $\xi^{\mu}$ and fluctuation 
$\eta^{\mu}$ are  condensed, that is, $m^{\mu}$ and $l^{\mu}$ are 
of order $ O(1)$ for a given $\mu=\nu$ (and for $\mu \neq \nu$ 
they are of order $O(1/\sqrt{N})$), and we call these $m$ and $l$, 
respectively. These yield a finite signal term for each local field 
and related Gaussian noise terms. In accordance with this, we also 
assume that $n=n^{\nu}=O(1)$ and $s=s^{\nu}=O(1)$ and we denote 
$i=i^{\nu}$. 

The asymptotic macrodynamics for the extremely diluted network follows 
then the single-step evolution equations for the order parameters, exact 
in the large-N limit, for each time step $t$ \cite{DK00},
\begin{eqnarray}
\label{19}
&&m_{t+1}= 
\int Dy\int Dz \,\,
F_{\beta}(\frac{m_t}{a}+y\Delta_t;
\frac{l_t}{a}+z\frac{\Delta_t}{1-a}),  \\
\label{20}
&&n_{t+1}=
\int Dy\int Dz \,\,
G_{\beta}(\frac{m_t}{a}+y\Delta_t;
\frac{l_t}{a}+z\frac{\Delta_t}{1-a}) ,  \\
\label{21}
&&s_{t+1}=
\int Dy\int Dz \,\,
G_{\beta}(y\Delta_t;-\frac{l_t}{1-a}+z\frac{\Delta_t}{1-a}),
\end{eqnarray}
together with the dynamic activity $q_{t}= a n_{t} + (1-a) s_{t}$.  
The equation for $l_{t}$ 
is obtained using the relation $l_{t}=(n_{t}-q_{t})/(1-a)$. Here, 
as usual, $Dx=\exp(-x^{2}/2)dx/\sqrt{2\pi}$ whereas
 ${\Delta_t}^2={\alpha}q_{t}/a^2$ and ${\Delta_t}^2/(1-a)^2$ are 
the variances in the Gaussian local fields $h_{i}$ and $\theta_{i}$, 
respectively. 
With these equations we also get the time evolution of the information 
$i$ by means of Eqs. (12) to (14).

The equations for the dynamics of the macroscopic order parameters 
can now be used to study both the time evolution of 
the network and to determine the properties of the stable stationary
states.

\section{Results}

We recognize, essentially, three phases given by stable stationary
states of the network dynamics Eqs. (18)-(20), as shown in 
Fig.1, for a typical activity of $a=0.8$.  There is a retrieval 
phase $R$ $(m\neq 0,l\neq 0)$, a fluctuation phase $Q$ $(m=0,l\neq 0)$ 
and a self-sustained activity phase $S$ $(m=0,l=0)$, referred to as 
the zero phase, $Z$, in previous works \cite{DK00,DK02}, all for 
$q\sim a$. The stationary states in these phases are indicated 
as attractors (a) in the side table of the phase diagram. 
There are also saddle-point solutions (s) either with 
$m=0$, $l\neq 0$, $q\sim a$ or with $m=0$, $l=0$, $q\sim a$, denoted 
also by $Q$ and $S$, respectively.  Both 
the saddle-points in $Q$ and $S$ have attractor directions along 
$l$, towards $l^{*}\neq 0$ and $l^{*}=0$, respectively, and 
repeller directions along $m$ away from $m=0$. Hence, they have 
strictly one-dimensional basins of attraction in the 
two-dimensional order parameter subspace of $m$ and $l$, and 
then only if the precise initial condition $m_{0}=0$ is met.

There is a retrieval phase in regions I to V. It is the only stable 
phase in regions I to III and it coexists with the self-sustained 
activity phase in regions IV and V. The latter implies that in
these two regions the basin of attraction of the retrieval states 
is always limited by the attracting self-sustained activity states. 
On the other hand, the fluctuation phase exists only in regions VI 
and VII, coexisting in the latter with the self-sustained activity 
phase which, in turn, exists only in that region and in region VIII. 
Dotted lines in the phase diagram denote continuous phase boundaries 
while full lines indicate discontinuous transitions. The phase 
boundaries denoted by thick lines mark the boundary of the
retrieval phase, the ones further to the right yield the critical storage 
capacity $\alpha_{c}$, where both overlaps disappear. Extensive calculations 
of the $\alpha$ dependence of the order parameters were performed to 
obtain the phase diagram, with particular emphasis in the search for 
possible stable $Q$ states. Clearly, it can be seen that, for a given 
activity $a$, these states appear only above a certain threshold in 
$T$.

A similar behavior appears for other big values of $a$ above a minimum, 
whereas a different behavior sets in for smaller $a$, as shown in Fig. 2 
for $T=T(a)$ and $\alpha=0$. Indeed, when $a$ is less than $1/2$ there 
is a continuous phase boundary at $T=2/3$ between the retrieval phase 
at low $T$ and the self-sustained activity phase at high $T$. At $a=1/2$ 
the transition becomes discontinuous and, up to $a=0.698$, the only phases 
present are still $R$ and $S$. The transition remains discontinuous 
for small $\alpha$ and it becomes continuous for bigger values of 
$\alpha$. For $\alpha=0$, the $Q$ phase starts to appear with increasing 
$a$ at a triple point with $a=0.698$ and $T=0.767$.  
Beyond $a=0.698$ the transition between the $R$ and the $Q$ phase
remains discontinuous up to the tricritical point with $a=0.711$ and  
$T=0.785$. For bigger values of $a$, the transition between the $R$ and 
the $Q$ phase remains continuous and ends at $T=1/2$ for $a=1$. It also 
turns out that there is a critical $\alpha=1/\pi\approx 0.318$, for $a=1$.

We discuss next the typical $\alpha$ dependence of the order parameters 
that yield the phase diagram of Fig. 1 and we also show the 
information content of the network below and above the threshold 
where the $Q$ phase starts 
to appear. For  $a=0.8$ where the threshold is given by $T\approx 0.45$
for $\alpha=0.221$, we 
obtain the results shown in Fig. 3. Clearly, for $T$ below that threshold 
(left figure) the two overlaps m and l remain finite together, in a behavior 
characteristic of retrieval, up to the critical $\alpha_{c}$. Thus, 
in this regime the fluctuation overlap does not yield anything 
essentially new that is not contained in the retrieval overlap. 
In contrast, above the threshold (right figure) the retrieval overlap 
disappears  first with increasing $\alpha$ leaving a finite $l\neq 0$ that 
describes a fluctuation overlap up to a bigger critical value 
$\alpha_{c}$. Hence, it is necessary that first  $T$ 
and then $\alpha$ become large enough for the $Q$ states to become 
stable. It is also worth noting that the fluctuation overlap carries a 
finite information even with zero pattern retrieval overlap. Thus, 
although the information transmitted by the network is mainly in the 
retrieval phase, there is also some information due to the $Q$ phase.
This information is provided by the fact that the active neurons 
coincide with the active patterns but the signs are not correlated. 
One might imagine an example in pattern recognition
where, looking at a black and white picture on a grey background, this
phase would tell us the exact location of the picture with respect to
the background without finding the details of the picture itself.

We also show in Fig. 3 the comparison of the performance with two other
three-state networks. One is the usual network with an externally 
adjustable optimal threshold parameter \cite{Y89,BS94} that 
appears in the quadratic part of the single-site energy function, 
formally the same as Eq.(\ref{4}) but with a uniform $\theta=\theta_i$.  
The parameter is restricted to be positive and chosen to yield the 
largest mutual information. Allowing the threshold to become negative 
would essentially mimick a binary model and this is not the subject of the 
present work. The second network is a phenomenological extension to 
finite $T$ of a recent three-state self-control model (SCT) \cite{BA00,BD00} 
in which the  self-control threshold $\theta_{t}$ at $T=0$ is replaced by a 
linearly shifted threshold $\overline{\theta_{t}}=\theta_{t}-T$, 
where $\theta_{t}=\sqrt{2\ln a}\,\, D_{t}$ with 
$D_{t}^{2}=\alpha q_{t}/a$ being the variance of the noise. The results of 
Fig. 3 clearly show that the BEG network has a better performance for 
high $T$, at least as far as the information content is concerned, than 
the optimal threshold network, and a worse performance for lower $T$.

To understand the typical behavior of  the dynamics of the network we 
show in Fig.4 the time evolution of the  order parameters and the 
information content, for $a=0.8$ and $T=0.6$, in both the BEG and 
the SCT  network. In support of the phase diagram  
shown in Fig. 1, it can be seen that in the case of the BEG network 
with increasing $\alpha$, one has first the asymptotic states of an 
$R$ phase, then the states of a $Q$ phase and, finally, the states of 
the $S$ phase. In all cases where one would expect a $Q$ state, we 
start with the most favorable initial overlaps for that state, that 
is $m_{0}=0$ and a small but finite $l$. In contrast, for the 
SCT network and the indicated values of $\alpha$ one has only an R 
phase.
 
A closer examination of the curves for the BEG network reveals that
the fluctuation overlap may ``drive'' a vanishingly small initial 
retrieval overlap, meaning almost no recognition of a given pattern by 
the state of the network, into an asymptotic state with finite recognition. 
This is in contrast 
with the expectation for other three-state networks, as in the case of 
the SCT network, where first the overlap $m_{t}$ becomes non-zero. It 
is also worth noting that, with a very small initial $m_{0}$, the states 
of the network are expected to pass through the vicinity of a saddle 
point, with a finite fluctuation overlap $l$ and still a
vanishing retrieval overlap at small or intermediate times. This 
situation is described by the first plateaus in $q$, $l$ and $i$. It is 
only in passing beyond those plateaus, which may take a rather
long time, that the states attain the asymptotic behavior of the 
retrieval phase. It also turned out that with the initial conditions used
for the BEG network, in the left part of the figure, there is no retrieval 
in the SCT network, meaning that the basins of attraction for retrieval 
are larger in the BEG network. 

Finally, the results for the stationary states are confirmed by a set of 
flow diagrams, a particular one is shown in Fig. 5 for $a=0.8$ and 
$T=0.6$, first below the $R$-$Q$ phase boundary, where the 
retrieval state $R$ is stable, and then above, where the $Q$ phase is 
stable. Clearly, in the first case, for a small initial retrieval 
overlap and a finite fluctuation overlap, the states evolve first in the
attractor direction of the saddle point and only then they start to 
flow towards the true (retrieval) attractor. Similar flow diagrams were 
obtained for other sets of parameters and in all cases we found that the 
attractors have fairly large basins of attraction.

\section{Conclusions}

We discussed the dynamic evolution and the stationary states of the 
recently introduced BEG neural network model for an extremely diluted 
architecture. We made particular emphasis on the stability of the 
stationary states, which had not been explored before, and found that 
the new phase $Q$ (called previously, the ``dipolar'' or
``quadrupolar'' phase), characterized 
by a zero retrieval overlap and a finite fluctuation overlap, is a true 
stable phase only for moderately to large pattern activity $a$. We 
found that an activity dependent synaptic noise has a relevant role in 
deciding whether the new phase can be reached or not. In particular, 
that phase is not a stable one at $T=0$ for any activity smaller than 
one. 
It is also not stable, in general, for somewhat higher values of $T$.

We also found that the dynamics may be slowed down due to the presence 
of saddle-point solutions in the equations that appear in large regions 
of the phase diagram, in particular in the retrieval phase and close to 
the critical phase boundary. 
Although the specific results obtained here 
are for the extremely diluted network, some of the features found 
may also appear in other architectures, 
for instance, in a layered network, 
and there is work in progress for that case \cite{BE02}. 
It would be 
interesting to study the time evolution of the BEG network also in 
other non-trivial dynamics.

\section*{Acknowledgments}

One of the authors (DB) wants to thank T. Verbeiren and J. Busquets
Blanco for critical discussions and the Fund for Scientific
Research-Flanders, Belgium, for financial support. DRDC acknowledges 
a Ram\'on y Cajal grant from the Spanish Ministry of Science (MCyT), 
and thanks the K.U.Leuven, Belgium, for a visiting grant.  
EK warmly thanks for hospitality the Abdus Salam International Center, 
for Theoretical Physics, Trieste, 
and is financially supported by a grant 
DGI (MCyT) BFM 2001-291-C02-01. 
The work of WKT is partially supported by the 
Conselho Nacional de Desenvolvimento Cient\'{\i}fico e Tecnol\'ogico 
(CNPq), Brazil, and the same author thanks the Funda\c{c}\~ao de Amparo 
\`a Pesquisa do Estado de Rio Grande do Sul (FAPERGS), Brazil, for a Visitor 
Scientist grant to the IF-UFRGS.


\end{multicols}
\newpage

\vspace{-1.cm}
\centerline{
\begin{tabular}{|c|cccccccc|}
\hline
 & I & II & III & IV & V & VI & VII & VIII \\
\hline
R & a & a & a & a & a & - & - & - \\
\hline
Q & s & s & - & - & s & a & a & - \\
\hline
S & - & s & s & a & a & - & a & a \\
\hline
\end{tabular}
}

\vspace{-4cm}
\begin{figure}[h]
\begin{center}
\epsfysize=12cm
\leavevmode
\epsfbox[1 1 700 700]{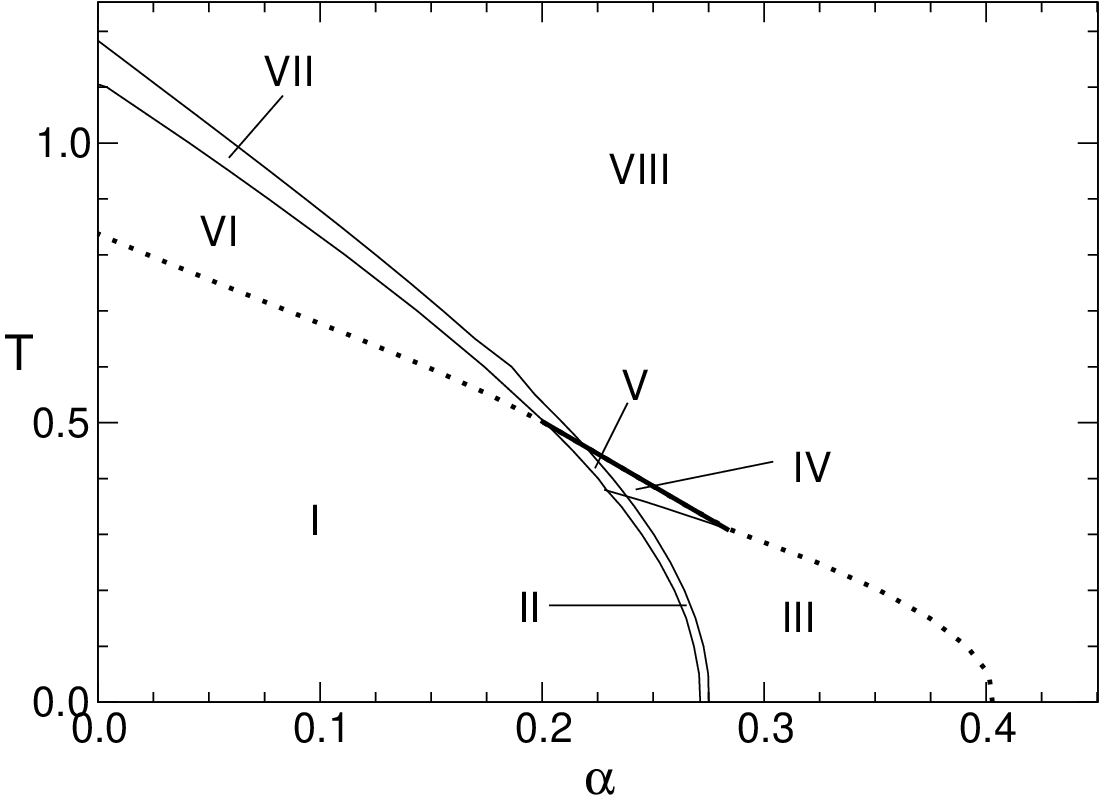}
\end{center}
\caption{ {\em 
The $(T,\alpha)$ phase diagram for the extremely diluted
BEG network with pattern activity $a=0.8$. 
Full (dotted) lines denote
discontinuous (continuous) transitions. 
The heavy lines denote the boundary of the retrieval phase $R$ 
and the other lines the boundaries of 
the fluctuation-overlap phase $Q$ and the self-sustained 
activity phase $S$. 
The solutions are either attractors ($a$) or saddle points ($s$)
.
} }
\label{f1}
\end{figure}


\vspace{-4cm}
\begin{figure}[h]
\begin{center}
\epsfysize=12cm
\leavevmode
\epsfbox[1 1 700 700]{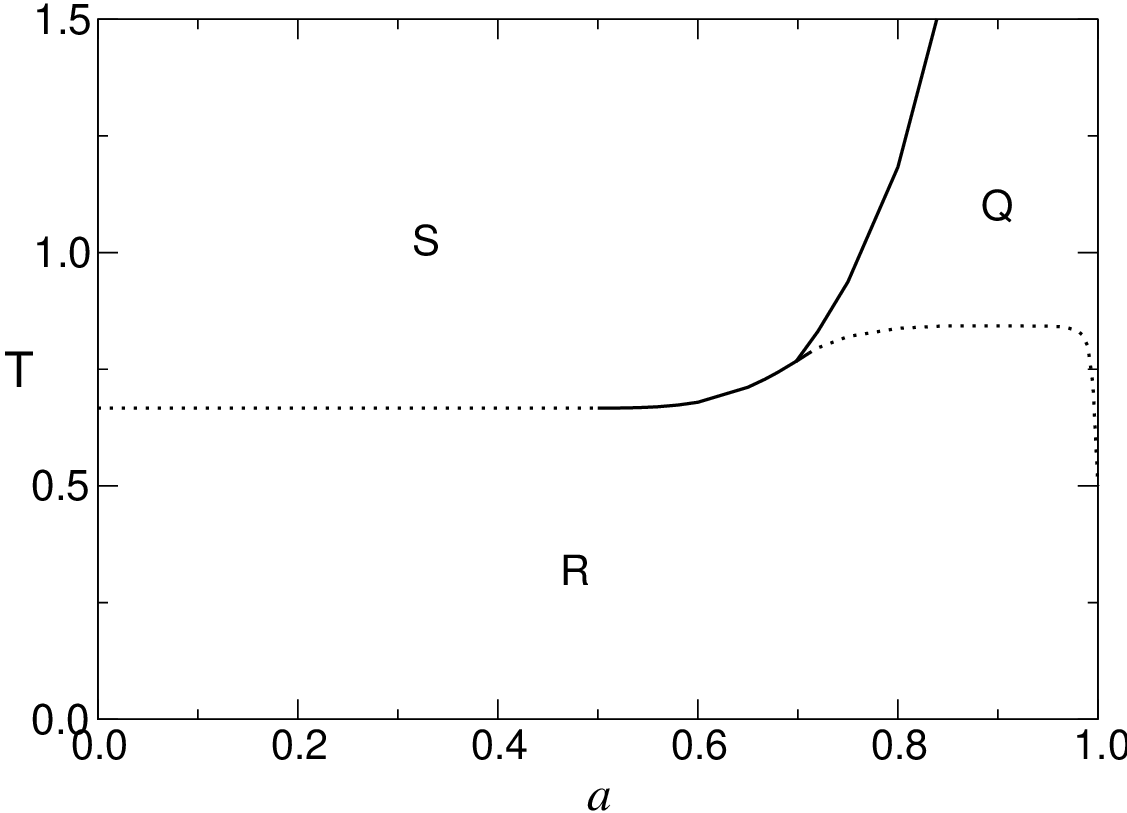}
\end{center}
\caption{ {\em 
The $(T,a)$ phase diagram for the extremely diluted BEG
network with the load $\alpha=0$. 
Full (dotted) lines denote discontinuous
(continuous) transitions
.
} }
\label{f2}
\end{figure}


\newpage

\vspace{-5cm}
\begin{figure}[h]
\begin{center}
\epsfysize=20cm
\leavevmode
\epsfbox[1 1 700 700]{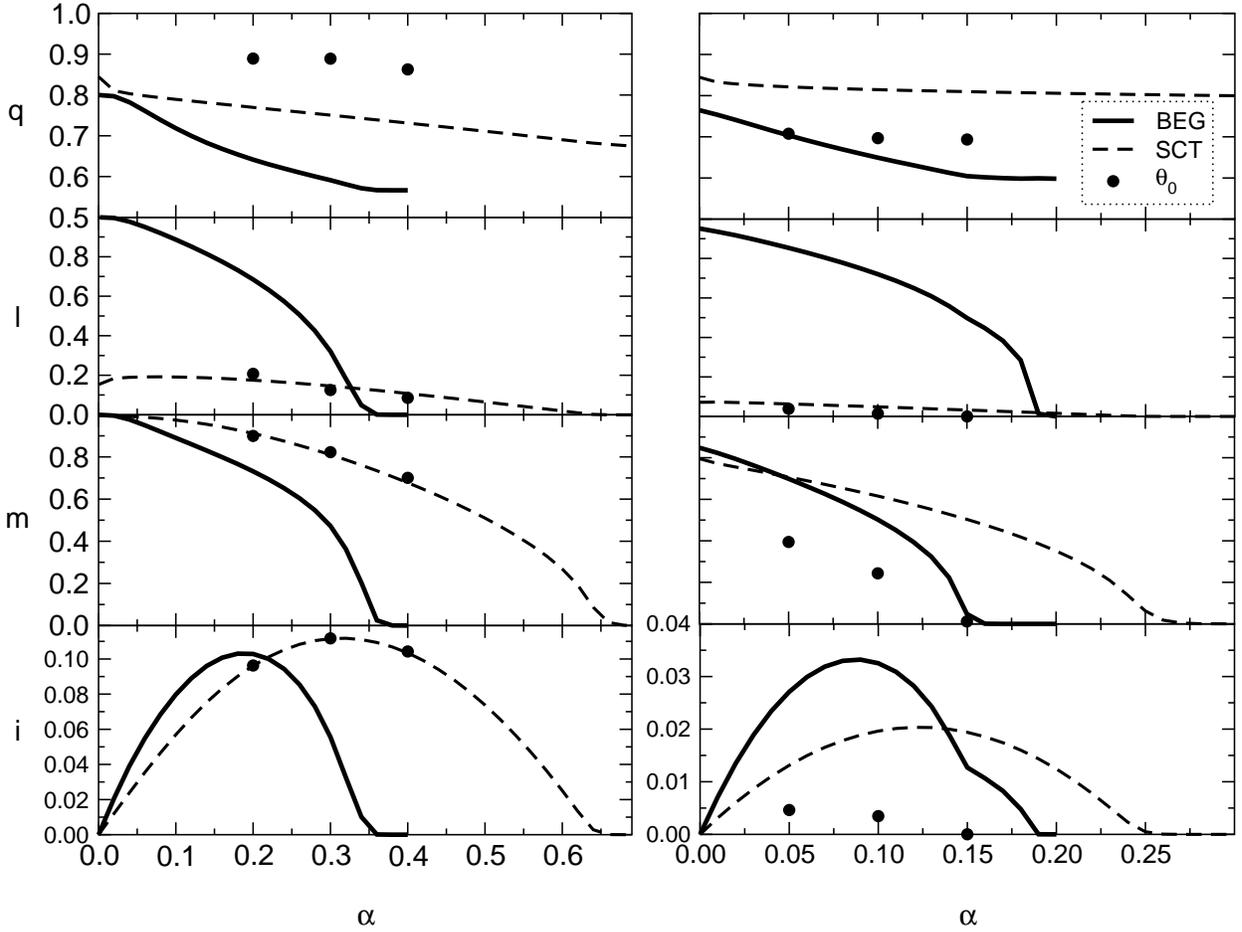}
\end{center}
\caption{ {\em 
The order parameters $m$, $l$, and $q$, and the
information  content $i$, in the stationary state, 
for initial overlaps $m_{0}=1$, $l_{0}=1$ and $q_{0}=a$, 
as functions of the load $\alpha$, 
for the BEG network with  pattern activity $a=0.8$ and two noise levels:
$T=0.2$ (left) and $T=0.6$ (right). 
The self-controlled threshold network (SCT), in dashed lines, 
and  the optimal threshold network (dots), are also shown
.
} }
\label{f3}
\end{figure}

\newpage

\vspace{-4cm}
\begin{figure}[h]
\begin{center}
\epsfysize=12cm
\leavevmode
\epsfbox[1 1 600 600]{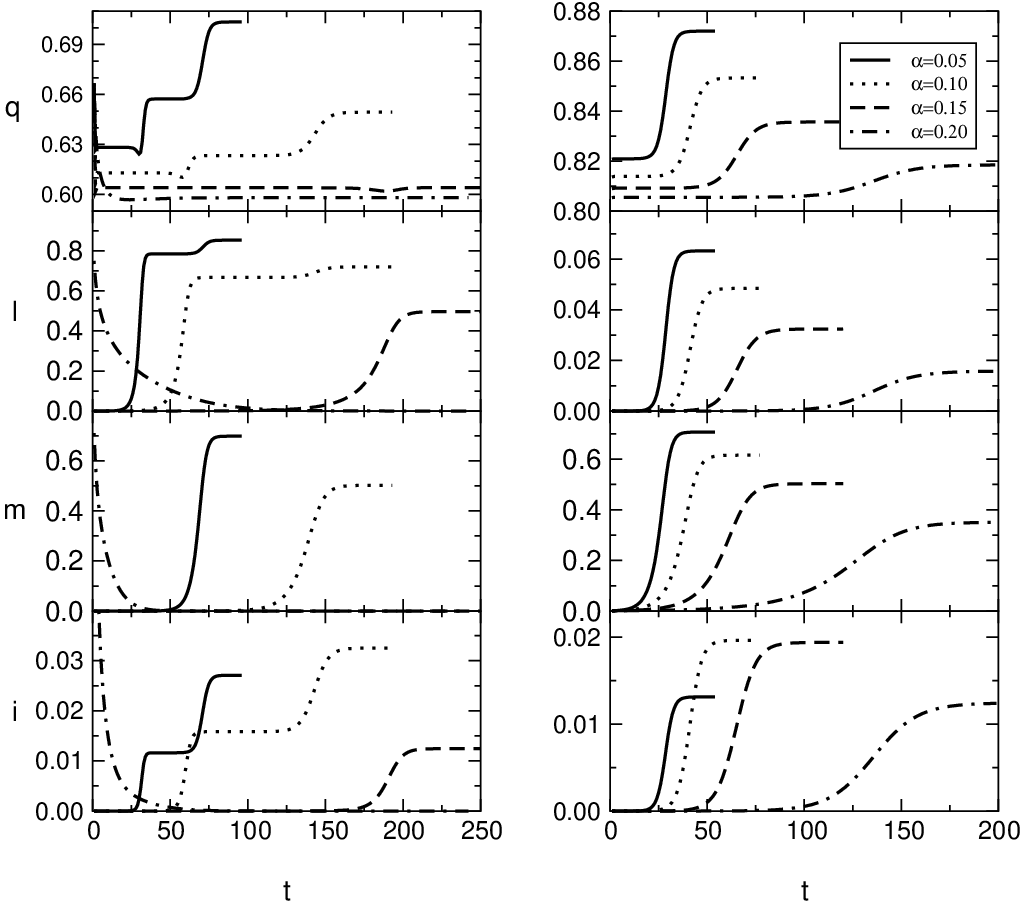}
\end{center}
\caption{ {\em 
Time evolution of the order parameters $m$, $l$ and $q$  
and information content $i$, for pattern activity $a=0.8$, 
temperature $T=0.6$ and load $\alpha$, as indicated. 
The BEG network (left), for the initial overlaps $m_{0}=l_{0}=10^{-5}$, 
except for $\alpha=0.2$ ($m_{0}=l_{0}=1$); 
the SCT network (right), for the initial overlaps $m_{0}=10^{-3}$ and 
$l_{0}=1$; both with $q_{0}=a$
.
} }
\label{f4}
\end{figure}


\vspace{-4cm}
\begin{figure}[h]
\begin{center}
\epsfysize=11.4cm
\leavevmode
\epsfbox[1 1 700 700]{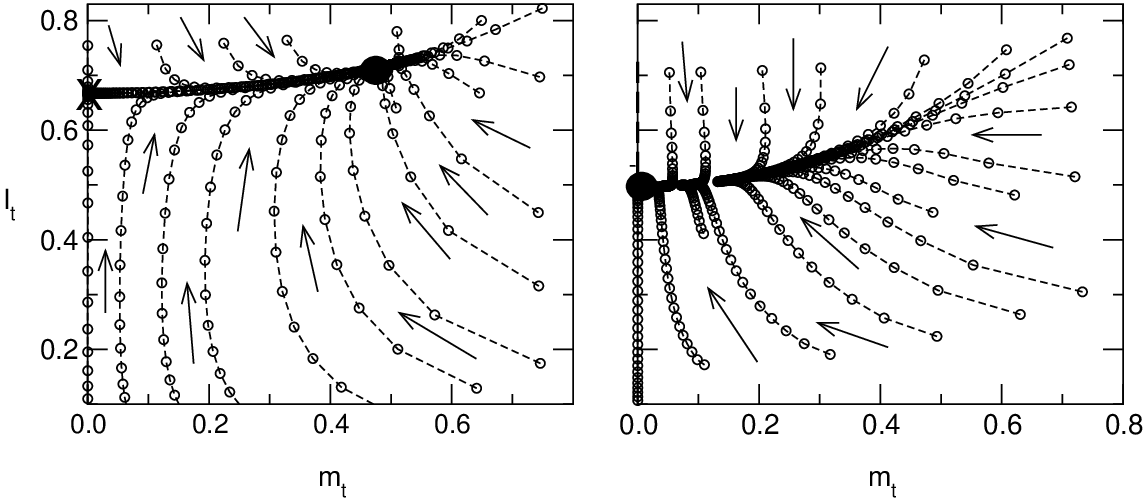}
\end{center}
\caption{ {\em 
The flow diagram $(l,m)$ for the BEG network with $a=0.8$ and $T=0.6$. 
For $\alpha=0.1$ (left) the stable attractor is $R$, indicated 
by the circle, while $Q$, indicated by the cross, is a saddle point. 
For $\alpha=0.15$ (right) the stable attractor is $Q$
.
} }
\label{f5}
\end{figure}


\end{document}